\documentclass[twocolumn,prb,superscriptaddress,noshowpacs,epsf]{revtex4}

\usepackage{graphicx}

\begin{document}

\title{Switchable coupling between charge and flux qubits}
\date{\today}
\author{Xiao-Ling He}
\affiliation{Department of Physics and Surface Physics Laboratory (National Key
Laboratory), Fudan University, Shanghai 200433, China}
\affiliation{Frontier Research System, The Institute of Physical and Chemical Research
(RIKEN), Wako-shi 351-0198, Japan}
\author{J. Q. You}
\affiliation{Department of Physics and Surface Physics Laboratory (National Key
Laboratory), Fudan University, Shanghai 200433, China}
\affiliation{Frontier Research System, The Institute of Physical and Chemical Research
(RIKEN), Wako-shi 351-0198, Japan}
\author{Yu-xi Liu}
\affiliation{Frontier Research System, The Institute of Physical and Chemical Research
(RIKEN), Wako-shi 351-0198, Japan}
\author{L. F. Wei}
\affiliation{Frontier Research System, The Institute of Physical and Chemical Research
(RIKEN), Wako-shi 351-0198, Japan}
\author{Franco Nori}
\affiliation{Frontier Research System, The Institute of Physical and Chemical Research
(RIKEN), Wako-shi 351-0198, Japan}
\affiliation{Center for Theoretical Physics, Physics Department, Center for the Study of
Complex Systems, The University of Michigan, Ann Arbor, MI 48109-1040, USA}

\begin{abstract}
We propose a hybrid quantum circuit with both charge and flux qubits 
connected to a large Josephson junction that gives rise to 
an effective inter-qubit coupling controlled by the external magnetic flux. 
This switchable inter-qubit coupling can be used to transfer back and forth 
an arbitrary superposition state between the charge qubit and the flux qubit 
working at the optimal point. The proposed hybrid circuit provides a promising 
quantum memory because the flux qubit at the optimal point can store the tranferred 
quantum state for a relatively long time.
\end{abstract}

\pacs{74.50.+r, 85.25.-j, 03.67.Lx}
\maketitle

\section{Introduction}

Charge and flux qubits are two different types of superconducting 
Josephson-junction (JJ) qubits for quantum 
computing (see, e.g., Ref.~\onlinecite{YN}). 
The charge qubit~\cite{NEC} has the advantage of more flexible controllability 
via external parameters; it can be conveniently controlled by 
the gate voltage and the applied magnetic flux. Namely, these parameters
control the longitudinal ($\sigma _{z}$) and transverse ($\sigma _{x}$) terms 
in the reduced Hamiltonian of the charge qubit. 
As for the flux qubit,~\cite{Orlando} 
the longitudinal term can be controlled by the applied magnetic flux, 
but it is hard to control the transverse term via an external parameter. 
In spite of this limitation, experiments~\cite{Bert} demonstrate that 
the flux qubit has a longer decoherence time than the charge qubit. 
Indeed, this is one of the major advantages for the flux qubit.

Here we propose a hybrid quantum circuit by connecting a large JJ to both charge and 
flux qubits. This large JJ serves as a bosonic data bus. By virtually exchanging 
bosons between the data bus and the qubits,
a $\sigma_x\sigma_z$-type interaction is produced between the charge and flux qubits.
Equivalently, this inter-qubit coupling is achieved as if the large JJ acts as 
an effective inductance. 
Indeed, charge qubits can be coupled by an inductance and the inductive  
coupling is switchable via either the applied magnetic flux~\cite{JTNPRL,MAK,JTNPRB}
or the current biasing the large JJ that acts as an effective nonlinear 
inductance.~\cite{JTNPRB,LAN} 
The advantage of this switchable inter-qubit coupling has been taken for proposing 
an efficient quantum computing.~\cite{JTNPRL}
Also, flux qubits can be coupled by an inductance,~\cite{JNN,JENA,MAJER} 
but the inter-qubit coupling is {\it not} switchable. 
To achieve controllable coupling between flux qubits, it was proposed to 
use variable-frequency magnetic fields applied to the qubits.~\cite{Devoret, LIU}  

The hybrid quantum circuit proposed here has the advantages of both charge 
and flux qubits. For instance, taking an  
advantage of the charge qubit, the coupling between the charge and flux qubits 
becomes switchable by varying the magnetic flux applied to the charge qubit.  
Also, it is easy to prepare an arbitrary superposition state for the charge qubit. 
Moreover, as we will show below, this arbitrary state can be conveniently transferred 
to the flux qubit using the controllable coupling between the charge and flux qubits. 
More importantly, in this case the flux qubit works at the optimal point 
and it has a relatively long decoherence time. This remarkable advantage of the flux 
qubit assures that the state transferred to the flux qubit can be stored for a 
longer time. Also, when needed, this stored state can be easily transferred back to 
the charge qubit. These features indicate that this hybrid circuit is suitable 
for a quantum memory.

\section{The Model}

The hybrid quantum circuit we consider is shown in Fig.~1, where a charge qubit and 
a flux qubit is coupled by a large JJ. The phase drops across the JJs in the two 
qubits are coupled to the phase drop $\gamma$ across the large JJ: 
\begin{eqnarray}
&&\varphi_1^{(1)}-\varphi_2^{(1)}-\gamma +2\pi f_1=0,  \nonumber\\
&&\varphi_1^{(2)}-\varphi_2^{(2)}+\varphi_3^{(2)}+\gamma + 2\pi f_2=0,
\end{eqnarray}
where the reduced magnetic fluxes are given by $f_i=\Phi_{ei}/\Phi_0$, with $i=1,2$. 
Here $\Phi_0$ is the magnetic flux quantum and $\Phi_{ei}$, $i=1,2$, are the 
magnetic fluxes applied to the charge and flux qubits, respectively. The Hamiltonian 
of the system is 
\begin{equation}
H=H_1+H_2+H_m. 
\end{equation}
Here $H_1$ is the Hamiltonian 
of the charge qubit in the presence of the (middle) large JJ:~\cite{JTNPRB}
\begin{equation}
H_1=E_{c1}(n-n_g)^2
-2E_{J1}\cos\left(\pi f_1-\frac{1}{2}\gamma\right)\cos\varphi,
\end{equation}
where $E_{c1}=e^2/C_J$ is the charging energy of the superconducting island 
in the charge qubit, $n=-i\partial/\partial\varphi$, $n_g=C_gV_g/2e$, 
and $\varphi=\frac{1}{2}(\varphi_1^{(1)}+\varphi_2^{(1)})$.

\begin{figure}
\includegraphics[width=3.4in,  
bbllx=59,bblly=565,bburx=460,bbury=740]{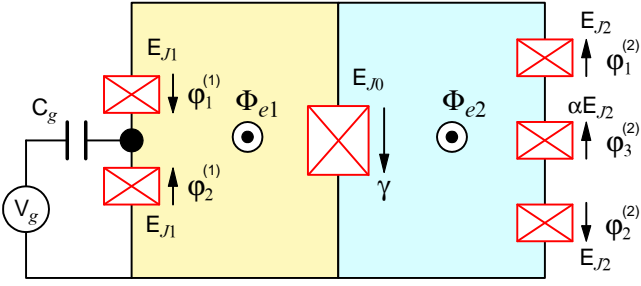} \caption{(Color
online) Schematic diagram of a hybrid quantum circuit with the charge (left) and 
flux (right) qubits connected to a large (middle) Josephson junction (JJ).
In the charge qubit, the two JJs have the same Josephson coupling energy $E_{J1}$ 
and capacitance $C_J$. In the flux qubits, two JJs have the same Josephson 
coupling energy $E_{J2}$ and capacitance $C_f$; the third JJ has a smaller Josephson 
coupling energy $\alpha E_{J2}$ and capacitance $\alpha C_f$, where $0.5<\alpha<1$. 
The large JJ in the middle has the Josephson coupling energy $E_{J0}$ and 
capacitance $C_0$. For the ideal hybrid circuit, $E_{J0}\gg E_{J2}\gg E_{J1}$.
Here the arrow near each JJ denotes the chosen direction for the positive phase 
drop across the corresponding junction.} 
\end{figure}

The Hamiltonian of the flux qubit in the presence of the large JJ is 
given by~\cite{Orlando} 
\begin{eqnarray}
H_2\!&\!=\!&\!\frac{P_p^2}{2M_p}+\frac{P_q^2}{2M_q}%
+2E_{J2}(1-\cos\varphi_p\cos\varphi_q)  \nonumber \\
&&+\alpha E_{J2}\left[1-\cos \left(2\varphi_q+\gamma + 
2\pi f_2\right)\right],
\end{eqnarray}%
where 
\begin{eqnarray}
P_k\!&\!=\!&\!-i\hbar{\partial}/{\partial\varphi_k}, \,\,\,\,k=p,q, \nonumber\\
M_p\!&\!=\!&\!2C_f\left({\Phi_0}/{2\pi}\right)^2, \\
M_q\!&\!=\!&\!M_p(1+2\alpha). \nonumber
\end{eqnarray}
The redefined phases $\varphi_p$ and $\varphi_q$ are 
$\varphi_p=\frac{1}{2}(\varphi_1^{(2)}+\varphi_2^{(2)})$
and $\varphi_q=\frac{1}{2}(\varphi_1^{(2)}-\varphi_2^{(2)})$.
The Hamiltonian of the middle large JJ reads
\begin{equation}
H_m=E_{c0}\left(N+\frac{1}{2}n_g\right)^2-E_{J0}\cos\gamma,
\end{equation}
where $E_{c0}=(2e)^2/2C_0$ is the charging energy of the large JJ and 
$N=-i\partial/\partial\gamma$.

We expand the potential in each qubit Hamiltonian into a series. 
When the middle JJ that couples the charge and flux qubits is large enough, 
one can retain the terms up to the first order of $\gamma$ 
(cf. Ref.~\onlinecite{JTNPRB}). 
The total Hamiltonian is then reduced to 
\begin{equation}
H=H_c+H_f+H_m+H_{\rm int},
\end{equation} 
where $H_c\equiv H_1(\gamma=0)$ is the Hamiltonian of 
the charge qubit without coupling to the large JJ and $H_f\equiv H_2(\gamma=0)$ 
is the Hamiltonian of the flux qubit without coupling to the large JJ.
The interaction Hamiltonian $H_{\rm int}$ bewteen the two qubits and the  
large JJ is 
\begin{equation}
H_{\rm int}=-\frac{\Phi_0}{2\pi}(I_1+I_2)\gamma, 
\label{INT1}
\end{equation}
where the circulating supercurrents in the (left) charge and (right) flux qubits
are given, respectively, by 
\begin{eqnarray}
I_1\!&\!=\!&\!I_{c1}\sin(\pi f_1)\cos\varphi, \nonumber\\
I_2\!&\!=\!&\!-\alpha I_{c2}\sin(2\varphi_q+2\pi f_2),
\end{eqnarray}
with $I_{ci}=2\pi E_{Ji}/\Phi_0$, $i=1,2$.  

When the charge states $|0\rangle$ and $|1\rangle$ are used as the basis states, 
the Hamiltonian $H_c$ can be reduced to (see, e.g., Refs.~\onlinecite{JTNPRL} 
and \onlinecite{MAK}) 
\begin{equation}
H_c=\varepsilon_1(V_g)\sigma_z^{(1)}-\Delta_1\sigma_x^{(1)}, 
\end{equation}
where 
\begin{eqnarray}
\varepsilon_1(V_g)\!&\!=\!&\!\frac{1}{2}E_{c1}(C_gV_g/e-1), \nonumber\\
\Delta_1\!&\!=\!&\!E_{J1}\cos(\pi f_1). 
\end{eqnarray}
The two charge states $|0\rangle$ and $|1\rangle$ 
correspond to zero and one extra Cooper pairs in the superconducting island, 
respectively. The circulating supercurrent $I_1$ is reduced to 
\begin{equation}
I_1=\frac{1}{2}I_{c1}\sin(\pi f_1)\sigma_x^{(1)}.
\label{I1}
\end{equation}  
Using the basis states $|\!\uparrow\rangle$ and $|\!\downarrow\rangle$ 
corresponding to the states with maximal clockwise and counter-clockwise persistent 
supercurrents in the flux qubit, one can reduce the Hamiltonian $H_f$ to 
(see, e.g., Ref.~\onlinecite{MAJER})
\begin{equation}
H_f=\varepsilon_2(\Phi_2)\sigma_z^{(2)}-\Delta_2\sigma_x^{(2)},
\end{equation} 
where 
\begin{equation}
\varepsilon_2(\Phi_2)=I_p\left(\Phi_2-\frac{1}{2}\Phi_0\right), 
\end{equation}
and $\Delta_2$ is the tunneling amplitude of the barrier in the double-well potential. 
The circulating supercurrent $I_2$ is reduced to 
\begin{equation}
I_2=I_p\sigma_z^{(2)},
\label{I2}
\end{equation}
where $I_p$ is the maximal persistent supercurrent of the flux qubit.

The middle JJ connecting the charge and flux qubits 
behaves like a particle with mass $M_0=2(\Phi_0/2\pi)^2C_0$, 
trapped in a cosinoidal potential $-E_{J0}\cos\gamma$.
Because this JJ is large, it can be approximately regarded as a harmonic oscillator:
\begin{equation}
H_m=\hbar\omega_p a^{\dagger}a,
\end{equation}  
with the plasma frequency 
\begin{equation}
\omega_p=\frac{\sqrt{8E_{J0}E_{c0}}}{\hbar}.
\end{equation} 
The boson operator is defined as 
\begin{equation}
a=(\xi/2)\gamma +i(1/2\xi)N, 
\end{equation}
with 
\begin{equation}
\xi=\left(\frac{E_{J0}}{2E_{c0}}\right)^{1/4}. 
\end{equation}
Thus, the phase drop $\gamma $ can be written as
\begin{equation}
\gamma =\frac{1}{\xi}\,(a^{\dagger}+a).
\label{gamma}
\end{equation}%
Substituting Eq.~(\ref{gamma}) into the interaction Hamiltonian (\ref{INT1}), 
one can write the 
total Hamiltonian of the hybrid circuit as  
\begin{equation}
H=H_0 + H_{\rm int}, 
\end{equation}
where
\begin{eqnarray}
H_0 \!&\!=\!&\! \varepsilon_1(V_g)\,\sigma_z^{(1)} - \Delta_1\sigma_x^{(1)} \nonumber\\
&&\!+\varepsilon_2(\Phi_2)\,\sigma_z^{(2)} - \Delta_2 \,\sigma_x^{(2)} 
+\hbar\omega_p \,a^{\dagger}a,
\end{eqnarray}
and
\begin{equation}
H_{\rm int}=-\left(g_{10}\,\sigma_x^{(1)}+g_{20}\,\sigma_z^{(2)}\right)
(a^{\dagger}+a),
\end{equation}
with
\begin{eqnarray}
g_{10}\!&\!=\!&\!\left({\Phi_0}/{2\pi}\right)
\left({I_{c1}}/{2\xi}\right)\sin(\pi f_1), \nonumber\\
g_{20}\!&\!=\!&\!\left({\Phi_0}/{2\pi}\right)\left({I_p}/{\xi}\right). 
\end{eqnarray}
This total Hamiltonian is analogous to 
two qubits separately coupled to an optical mode in a quantum cavity.

\section{Effective inter-qubit coupling}

Here we consider the case with the plasmon energy splitting $\hbar\omega_p$ 
much larger than the qubit energy splitting $\hbar\omega_q$. 
Now the rotation-wave approximation cannot be used since the condition
$\omega_p+\omega_q\gg \omega_p-\omega_q$, required for the 
rotation-wave approximation, is not satisfied here. Below we show that an effective 
interaction between the two qubits can be generated. Actually, because the plasmon 
energy splitting is much larger than the energy splittings of the qubits, 
the harmonic oscillator can be assumed 
to remain in the ground state, irrespective of the coupling of the large JJ to 
the qubits. Therefore, an effective inter-qubit interaction
is achieved by exchanging virtual bosons between the large JJ and the qubits. 

In the interaction picture, the system evolves as 
\begin{equation}
|\Psi(t)\rangle=U(t,0)|\Psi(0)\rangle, 
\end{equation}
where the evolution operator, up to  
second order, reads~\cite{quantum}
\begin{eqnarray}
U(t,0)\!&\!=\!&\! 1+\frac{1}{i\hbar}\int_0^tV_I(t_1)dt_1 \nonumber\\
&&\!+\left(\frac{1}{i\hbar}\right)^2\int_0^t dt_1\int_0^{t_1}V_I(t_1)V_I(t_2)dt_2 \,.
\label{pert}
\end{eqnarray}
The interaction operator $V_I(t)$ is 
\begin{equation}
V_I(t)=U_0^{\dag}(t)H_{\rm int}U_0(t)
\end{equation}
with $U_0(t)=\exp(-iH_0t)$. 

Here we use the basis states: 
\begin{eqnarray}
&&|\Psi_{1n}\rangle\equiv |g^{(1)},g^{(2)} \rangle \otimes |n\rangle, \nonumber\\
&&|\Psi_{2n}\rangle\equiv |g^{(1)},e^{(2)}\rangle \otimes |n\rangle, \nonumber\\
&&|\Psi_{3n}\rangle\equiv |e^{(1)},g^{(2)}\rangle \otimes |n\rangle, \nonumber\\
&&|\Psi_{4n}\rangle\equiv |e^{(1)},e^{(2)}\rangle \otimes |n\rangle, \nonumber
\end{eqnarray}
where $|g^{(i)}\rangle$ and $|e^{(i)}\rangle$ are the ground and excited states 
of the qubit $i$ ($i=1,2$), and $|n\rangle$ corresponds to the state of the harmonic 
oscillator with $n$ bosons. Equation (\ref{pert}) can be written as
\begin{eqnarray}
U(t,0)\!&\!=\!&\! 1+\frac{1}{i\hbar}\int_0^tF(t_1)dt_1 \nonumber\\
&&\!+\left(\frac{1}{i\hbar}\right)^2\int_0^t dt_1\int_0^{t_1}G(t_1,t_2)dt_2 \,,
\end{eqnarray}
where 
\begin{equation}
F(t_1)=\sum_{i,m}|\Psi_{im}\rangle\langle\Psi_{im}|V_I(t_1)
\sum_{j,n}|\Psi_{jn}\rangle\langle\Psi_{jn}| \,,
\end{equation}
and
\begin{eqnarray}
G(t_1,t_2)\!&\!=\!&\!\sum_{i,m}|\Psi_{im}\rangle\langle\Psi_{im}|V_I(t_1)
\sum_{j,n}|\Psi_{jn}\rangle\langle\Psi_{jn}| \nonumber\\
&&\times V_I(t_2)\sum_{k,p}|\Psi_{kp}\rangle\langle\Psi_{kp}| \,,
\end{eqnarray}
with $i,j,k=1,2,3$ and 4, and $m,n,p=0,1,2,\dots$.
After tedious calculations and neglecting the fast oscillatory terms, we have
\begin{equation}
U(t,0) \approx  1+\frac{1}{i\hbar}\int_0^t V_{\rm eff}\,dt_1 \,,
\end{equation}
where
\begin{equation}
V_{\rm eff}= \chi \,\sigma_x^{(1)}\sigma_z^{(2)}, 
\label{eff}
\end{equation}
with
\begin{equation}
\chi=\frac{2g_{10}g_{20}}{\hbar\omega_p}
=\left(\frac{\Phi_0}{2\pi}\right)^2\frac{I_{c1}I_p\sin(\pi f_1)}{2E_{J0}}.
\end{equation}
The reduced interaction Hamiltonian $V_{\rm eff}$ 
corresponds to an effective coupling between the two qubits after eliminating the 
degree of freedom of the large JJ.

Converted to the Sch{\"o}dinger picture and neglecting the fast oscillatory terms, 
the total Hamiltonian is then reduced to
\begin{eqnarray}
H \!&\!=\!&\! \varepsilon_1(V_g)\,\sigma_z^{(1)}
-\Delta_1\sigma_x^{(1)} \nonumber\\
&&\! +\varepsilon_2(\Phi_2)\,\sigma_z^{(2)}
-\Delta_2 \,\sigma_x^{(2)} + \chi \,\sigma_x^{(1)}\sigma_z^{(2)}.
\label{total}
\end{eqnarray}
It is interesting to note that the resulting inter-qubit coupling 
$\chi \,\sigma_x^{(1)}\sigma_z^{(2)}$ implies that the large JJ can behave like 
an inductance of value 
\begin{equation}
L_J=\frac{\Phi_0}{2\pi I_{c0}},
\end{equation} 
where $I_{c0}=2\pi E_{J0}/\Phi_0$. To see this, one can directly 
use the expressions, Eqs.~(\ref{I1}) and (\ref{I2}), of the circulating 
supercurrents $I_i$, $i=1,2$, 
for charge and flux qubits and calculate the inductive inter-qubit coupling 
using the inductance $L_J$:
\begin{eqnarray}
L_J I_1I_2 \!&\!=\!&\! \frac{1}{2}L_J I_{c1}I_p\sin(\pi f_1)
\sigma_x^{(1)}\sigma_z^{(2)} \nonumber\\
&\equiv \!&\! \chi \sigma_x^{(1)}\sigma_z^{(2)}.
\end{eqnarray}
This is just the inter-qubit coupling given in 
Eqs.~(\ref{eff})-(\ref{total}). In particular,
when $\Phi_{e1}=0$, the charge qubit has no loop current and the coupling
between the charge and flux qubits is switched off. 

\section{Quantum memory}

Below we show a typical two-qubit gate achieved using Hamiltonian (\ref{total}).
This gate is called iSWAP and can be conveniently used to transfer an arbitrary 
unknown state of the charge qubit to the flux qubit working at the optimal 
point.
 
Let $\Phi_{ei}=\frac{1}{2}\Phi_0$, with $i=1,2$, 
so that $\Delta_1(\Phi_{e1})=\varepsilon_2(\Phi_{e2})=0$. Moreover, we choose 
a suitable gate voltage $V_g$ to have $\varepsilon_1(V_g)=-\Delta_2\equiv -\Delta$,
The Hamiltonian (\ref{total}) is reduced to 
\begin{equation}
H=-\Delta\sigma _{z}^{(1)}-\Delta\sigma_{x}^{(2)}
+\chi \sigma_{x}^{(1)}\sigma_{z}^{(2)}.  
\end{equation}
With the basis states $|0g\rangle$, $|1g\rangle$, $|0e\rangle$ and $|1e\rangle$, 
the two-qubit evolution operator $U=\exp(-iHt/\hbar)$ can be written as 
\begin{equation}
U=\left(
\begin{array}{cccc}
a & 0 & 0 & d \\ 
0 & b & c & 0 \\ 
0 & c & b & 0 \\ 
d & 0 & 0 & a^{*}
\end{array}
\right), 
\end{equation}
where
\begin{eqnarray} 
a\!&\!=\!&\!\cos(\Omega t)-i\frac{2\Delta\sin(\Omega t)}{\Omega}, \nonumber\\
b\!&\!=\!&\!\cos(\chi t), \nonumber\\ 
c\!&\!=\!&\!-i\sin (\chi t), \nonumber\\ 
d\!&\!=\!&\!-i\frac{2\chi\sin(\Omega t)}{\Omega}, 
\end{eqnarray}
with $\Omega =(4\Delta ^{2}+\chi ^{2})^{1/2}$. 

When $\cos(\Omega\tau)=1$ and $\sin(\chi\tau)=-1$, the two-qubit operation is an 
iSWAP gate:
\begin{equation}
U_{\rm iSWAP}=\left( 
\begin{array}{cccc}
1 & 0 & 0 & 0 \\ 
0 & 0 & i & 0 \\ 
0 & i & 0 & 0 \\ 
0 & 0 & 0 & 1
\end{array}
\right). 
\end{equation}
This gate can be achieved by choosing $\Omega\tau=2n\pi$ and 
$\chi\tau=(2m-\frac{1}{2})\pi$, which requires that
\begin{equation}
\frac{\chi}{\Omega}=\frac{4m-1}{4n},
\label{cond}
\end{equation}
where $m,n=1,2,3,\dots$. For instance, $\chi/\Omega=1/8$ when $m=1$ and $n=6$,
which gives $\chi/\Delta\approx 0.25$.

In order to transfer an arbitrary superposition state of the charge qubit to the flux 
qubit, we first prepare the flux qubit at the ground state $|g\rangle$ and then 
apply the iSWAP gate to the two-qubit system. This gives rise to
\begin{equation}
(\alpha |0\rangle +\beta |1\rangle )|g\rangle 
\Longrightarrow |0\rangle (\alpha |g\rangle +i\beta |e\rangle).
\end{equation}
Furthermore, to convert $i\beta |e\rangle$ to $\beta|e\rangle$, one can 
just freely evolve the flux qubit for a time $t=\pi\hbar/4\Delta_2$ 
after the interaction with the charge qubit is switched off (by choosing 
$\Phi_{e1}=0$). This corresponds to applying a one-qubit rotation  
$\exp \left(i\Delta_2\sigma_{x}^{(2)} t/\hbar \right)$ on the flux qubit 
for the time $t=\pi\hbar/4\Delta_2$. After this free evolution 
of the flux qubit, one has
\begin{equation}
\alpha |g\rangle +i\beta |e\rangle \Longrightarrow (\alpha |g\rangle
+\beta |e\rangle),
\end{equation}
up to a global phase factor that produces no effect on the quantum state.
Therefore, an arbitrary unknown state $\alpha|0\rangle + \beta|1\rangle$ 
of the charge qubit is finally transferred to the flux qubit as 
$\alpha|g\rangle + \beta|e\rangle$. 
Because the flux qubit works at the optimal point and it has a relatively long 
decoherence time, the above operations provide a promising way to achieve 
a quantum memory that stores the quantum information for a longer time in 
the flux qubit. Also, when needed, the quantum information stored in the flux 
qubit can be converted back to the charge qubit by just successively applying
the above operations in the reverse manner. 

Finally, we estimate the coupling strength between the charge and flux qubits.
For instance, one can choose $E_{J0}=5E_{J2}$. Typically, the flux qubit has an 
energy splitting $\Delta_2\approx 0.02E_{J2}$ when $\alpha \approx 0.75$ 
and the maximal persistent supercurrent is $I_p\sim 0.5I_{c2}=\pi E_{J2}/\Phi_0$.
If the Josephson coupling energies are chosen as $E_{J1}=0.04E_{J2}$ for both 
charge and flux qubits, the inter-qubit coupling strength is $\chi\sim 0.002E_{J2}$;
when $E_{J1}=0.1E_{J2}$, $\chi\sim 0.005E_{J2}$. Thus, one has $\chi$ in the 
range of $0.6$~GHz $\sim 1.5$~GHz for a typical Josephson coupling energy 
$E_{J2}=300$~GHz. This inter-qubit coupling strength is strong enough for 
achieving a fast two-qubit operation used for the quantum memory.
Moreover, for $\chi\sim 0.005E_{J2}$ and $\Delta\equiv \Delta_2\approx 0.02E_{J2}$,
one has $\chi/\Delta\sim 0.25$, which corresponds to Eq.~(\ref{cond}) 
with $m=1$ and $n=6$. This implies that the iSWAP gate is achievable by 
choosing suitable parameters for the hybrid quantum circuit. 

\section{Conclusion}

In conclusion, we have proposed a hybrid quantum circuit where both charge and flux 
qubits are connected to a large JJ. This large JJ gives rise to 
an effective inductive coupling between the charge and flux qubits, 
which is switchable via the magnetic flux applied 
to the charge qubit. Moreover, the resulting inter-qubit coupling  
can be used to transfer an arbitrary superposition 
state of the charge qubit to the flux qubit working at the optimal point. 
This hybrid circuit provides a promising quantum memory because the flux qubit 
at the optimal point can store the tranferred quantum state for a long 
time.

\begin{acknowledgments}
This work was supported in part by the NSA, LPS and ARO.
X.L.H. and J.Q.Y. were supported by the SRFDP, 
the NFRPC grant No.~2006CB921205 
and the National Natural Science 
Foundation of China grant Nos.~10534060 and 10625416.
\end{acknowledgments}


\begin{thebibliography}{*}
\bibitem{YN} J.Q. You and F. Nori, Phys. Today {\bf 58}(11), 42 (2005), 
and references therein.

\bibitem{NEC} Y. Nakamura, Yu.A. Pashkin, and J.S. Tsai, Nature (London) 
{\bf 398}, 786 (1999).

\bibitem{Orlando} 
T.P. Orlando, J.E. Mooij, L. Tian, C.H. van der Wal, L.S. Levitov, S. Lloyd, and 
J.J. Mazo, Phys. Rev. B {\bf 60}, 15398 (1999).

\bibitem{Bert} P. Bertet, I. Chiorescu, G. Burkard, K. Semba, C.J.P.M. Harmans, 
D.P. DiVincenzo, and J.E. Mooij, Phys. Rev. Lett. {\bf 95}, 257002 (2005).

\bibitem{JTNPRL} J. Q. You, J. S. Tsai, and F. Nori, Phys. Rev. Lett. 
{\bf 89}, 197902 (2002); see also {\it New Directions in Mesoscopic Physics}, 
edited by R. Fazio, V.F. Gantmakher and Y. Imry (Kluwer, Dordrecht, 2003), 
pp. 351-360

\bibitem{MAK} Y. Makhlin, G. Sch{\"o}n, and A. Shnirman, Nature (London)
{\bf 398}, 305 (1999).

\bibitem{JTNPRB} J.Q. You, J.S. Tsai, and F. Nori, Phys. Rev. B {\bf 68}, 
024510 (2003).

\bibitem{LAN} J. Lantz, M. Wallquist, V.S. Shumeiko, and G. Wendin,
Phys. Rev. B {\bf 70}, 140507(R) (2004).

\bibitem{JNN} J.Q. You, Y. Nakamura, and F. Nori, Phys. Rev. B {\bf 71}, 
024532 (2005); 
B.L.T. Plourde, J. Zhang, K.B. Whaley, F.K. Wilhelm, T.L. Robertson, T. Hime, 
S. Linzen, P.A. Reichardt, C.-E. Wu, and J. Clarke, Phys. Rev. B {\bf 70}, 
140501 (2004).

\bibitem{JENA} A. Izmalkov, M. Grajcar, E. Il'ichev, Th. Wagner, H.-G. Meyer,  
A.Yu. Smirnov, M.H.S. Amin, A. Maassen van den Brink, and A.M. Zagoskin, 
Phys. Rev. Lett. {\bf 93}, 037003 (2004).

\bibitem{MAJER} J.B. Majer, F.G. Paauw, A.C.J. ter Haar, C.J.P.M. Harmans, 
and J.E. Mooij, Phys. Rev. Lett. {\bf 94}, 090501 (2005).

\bibitem{Devoret} C. Rigetti, A. Blais, and M. Devoret, Phys. Rev. Lett. 
{\bf 94}, 240502 (2005).

\bibitem{LIU} Y.X. Liu, L.F. Wei, J.S. Tsai, and F. Nori, Phys. Rev. Lett. 
{\bf 96}, 067003 (2006); 
P. Bertet, C.J.P.M. Harmans, and J.E. Mooij, Phys. Rev. B {\bf 73}, 064512 (2006);
M. Grajcar, Y.X. Liu, F. Nori, and A.M. Zagoskin, Phys. Rev. B {\bf 74}, 
172505 (2006). 
 
\bibitem{quantum} See, e.g., E. Merzbacher, {\it Quantum Mechanics}, 3rd Ed. 
(John Wiley, New York, 1998), Chapt.~19;
Y.X. Liu, L.F. Wei, and F. Nori, Phys. Rev. A {\bf 72}, 033818 (2005).

\end{thebibliography}
\end{document}